\documentclass[10pt,prc,aps]{revtex4}
\draft
\usepackage{graphicx}
\usepackage{color}
\usepackage{mathtools}
\begin{document}

\title{Analysis of the neutron matter equation of state and the symmetry energy up to fourth order of chiral effective field theory}       
\author{            
Francesca Sammarruca  and Randy Millerson  }                                                           
\affiliation{ Physics Department, University of Idaho, Moscow, ID 83844-0903, U.S.A. 
}
\date{\today} 
\begin{abstract}
We present predictions for the neutron matter equation of state, from leading to fourth order of chiral effective field theory, using recently developed, accurate chiral nucleon-nucleon potentials. We find the impact of subleading three-neutron forces to be mild and attractive.
We also show order-by-order predictions for the symmetry energy, and discuss its density dependence in relation to empirical constraints.

\end{abstract}
\maketitle 
        
\section{Introduction} 
\label{Intro} 

The idealized, infinite system known as nuclear matter has traditionally served as a convenient testing ground for theories of nuclear forces. Symmetric nuclear matter (SNM) offers the opportunity to explore the relationship between saturation properties and bulk properties of finite nuclei, an issue that is receiving considerable attention~\cite{SM20,Hoppe19,Huether+2020,Simonis+17,Mor18,Atk+}.

  Theoretical predictions of neutron-rich matter are equally important  and particularly timely, as they complement on-going and planned experimental efforts. PREX-II~\cite{S_etal_1_2011} has just been completed, and more experiments, such as CREX~\cite{MMMPRS_1_2013} and MESA~\cite{B_etal_1_2018}, seek to place high-precision constraints on the neutron radii and neutron skins of $^{48}$Ca and $^{208}$Pb, and will further elucidate the relation between the neutron skin and the energy and pressure in neutron-rich matter.

Furthermore, the equation of state (EoS) of neutron-rich matter has recently been brought to the forefront of nuclear astrophysics due to its relevance for calculating properties of neutron stars. Neutron stars are important natural laboratories for constraining theories of the EoS for neutron-rich matter, to which the mass-radius relationship of these stellar objects is sensitive. The radius of the average-mass neutron star is especially sensitive to the pressure gradient in neutron matter around normal densities~\cite{SM19}. Interest in these compact stars has increased considerably as we have entered the ``multi-messenger era" of astrophysical observation. The GW170817 neutron star merger event has yielded new and independent constraints on the radius of the canonical mass neutron star~\cite{A_etal_1_2017, A_etal_2_2017}.

It is the purpose of this paper to present our latest results for the EoS of neutron matter (NM) and the properties of the symmetry energy.
We will use high-quality nuclear forces constructed within the framework of chiral effective field theory (EFT). 
Over the past several years, chiral EFT has evolved into the most favorable approach  for developing nuclear interactions: it provides a systematic way to construct nuclear two- and 
many-body forces on an equal footing~\cite{Wei92} and allows
to assess theoretical uncertainties through an expansion controlled by an organizational scheme known as
``power counting"~\cite{Wei90}. Furthermore, chiral EFT maintains consistency with the underlying fundamental theory of strong interactions, quantum chromodynamics (QCD), through the symmetries and
symmetry breaking mechanisms of low-energy QCD.

We present  order-by-order calculations, including subleading three-nucleon forces (3NFs) up to next-to-next-to-next-to-leading order (N$^3$LO), with proper uncertainty quantification. Thus, we are able to draw reliable conclusions about the convergence pattern of the chiral perturbation series up to fourth order. 

Our main focal point is the symmetry energy.
It is puzzling that the findings from the recent PREX-II experiment~\cite{prexII} are at variance with a very large number of experimental measurements and theoretical predictions of the symmetry energy and its density dependence -- as noted by the authors of Ref.~\cite{prexII}. Therefore, we will also discuss our predictions in relation to empirical constraints. In particular, we provide microscopic values for the symmetry energy and its slope at densities at and below saturation density, in light of recent observations about the methods to extract those constraints.
 By constraining the symmetry energy over a range of densities below saturation -- having identified the densities at which a specific observable is maximally sensitive to the symmetry energy --  the model dependence arising from extrapolation to normal density can be avoided~\cite{Tsang}.

This paper is organized as follows. First, we review the main aspects of our calculations of NM and the symmetry energy, see Sec.~\ref{calc}. Predictions are presented and discussed in Sec.~\ref{res}, and
our observations and conclusions are summarized in Sec.~\ref{Concl}. 

\section{Description of the calculations} 
\label{calc}

 We perform microscopic calculations of the energy per particle in neutron matter using the nonperturbative particle-particle ladder approximation, which generates the leading-order contributions in the traditional hole-line expansion. 
The input two-nucleon forces (2NFs) and 3NFs are described next.
 
\subsection{The two-nucleon force}  
\label{II}

\begin{figure*}[!t] 
\centering
\hspace*{-1cm}
\includegraphics[width=9.0cm]{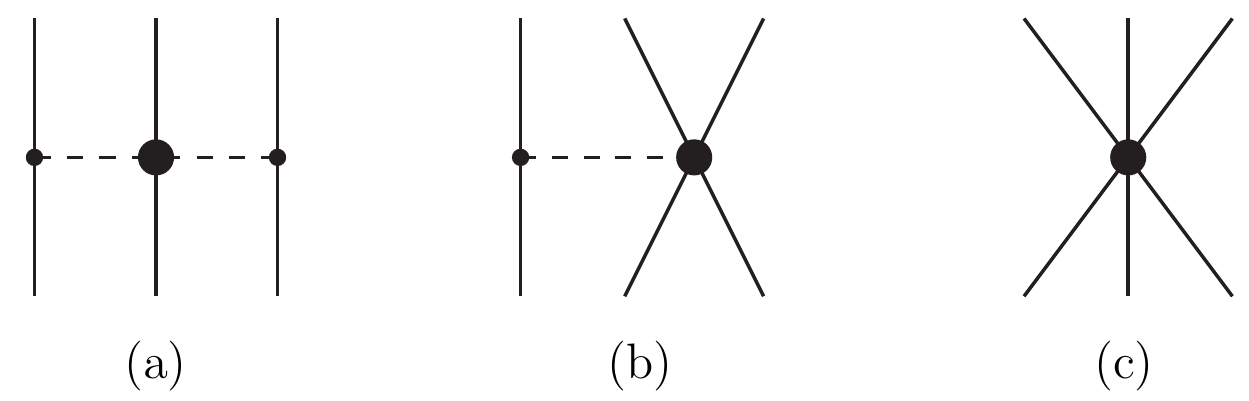}\hspace{0.01in} 
\vspace*{-0.05cm}
 \caption{ Diagrams of the leading 3NF: (a) the long-range 2PE, depending on the LECs $c_{1,3,4}$; (b) the medium-range 1PE, depending on the LEC $c_{D}$; and (c) the short-range contact, depending on the LEC $c_E$.}
\label{3nf_n2lo}
\end{figure*}

The 2NFs employed in this work are part of a set which spans five orders in the chiral EFT expansion, from leading order (LO) to fifth order (N$^4$LO)~\cite{EMN17}. For the construction of these potentials, the same power counting scheme and regularization procedures are applied through all orders, making this set of interactions more consistent than previous ones~\cite{EM03}.  Another novel and important aspect in the construction of these new potentials is the fact that the long-range part of the interaction is fixed by the $\pi N$ low-energy constants (LECs) as determined in the recent and very accurate analysis of Ref.~\cite{Hofe+}. In fact, for all practical purposes, errors in the $\pi N$ LECs are no longer an issue with regard to uncertainty quantification, which we will estimate based on chiral truncation error.
Furthemore, at the fifth (and highest) order, the nucleon-nucleon ($NN$) data below pion production threshold are reproduced with excellent precision ($\chi ^2$/datum = 1.15). (Of course, we will use the neutron-neutron versions of the potentials.)

Iteration of the potential in the Lippmann-Schwinger equation requires cutting off high-momentum components, consistent with the fact that chiral perturbation theory amounts to building a low-momentum expansion. This is accomplished through the application of a regulator function for which the non-local form is chosen:
\begin{equation}
f(p',p) = \exp[-(p'/\Lambda)^{2n} - (p/\Lambda)^{2n}] \,,
\label{reg}
\end{equation}
where $p' \equiv |{\vec p}\,'|$ and $p \equiv |\vec p \, |$ denote the final and initial nucleon momenta in the two-nucleon center-of-mass system, respectively. We use 
 $\Lambda$ = 450 MeV throughout this work.
The potentials are relatively soft as confirmed by the 
Weinberg eigenvalue analysis of Ref.~\cite{Hop17} and in the context of the perturbative calculations of infinite matter of  Ref.~\cite{DHS19}.

\begin{table*}[t]
\caption{Values of the LECs used in this work. 
The LECs $c_{1,3,4}$ are given in units of GeV$^{-1}$, and $C_S$ and $C_T$ are in units of 10$^4$ GeV$^{-2}$. $n$ refers to the exponent of the regulator function, Eq.~(\ref{reg}) ($\Lambda$=450 MeV).} 
\label{lecs_I}
\begin{tabular*}{\textwidth}{@{\extracolsep{\fill}}ccccccc}
\hline
\hline
 Order & $n$ & $c_1$ & $c_3$ & $c_4$   & $C_S$  & $C_T$ \\
\hline    
\hline
N$^2$LO  & 2& --0.74 & --3.61 & 2.44    &         &           \\
N$^3$LO  & 2& --1.07 & --5.32 & 3.56      & --0.011813 & --0.000025                                \\ 
\hline
\hline
\end{tabular*}
\end{table*}

\subsection{The three-nucleon force} 
\label{III} 

\begin{figure*}[!t] 
\centering
\hspace*{-1cm}
\includegraphics[width=11.5cm]{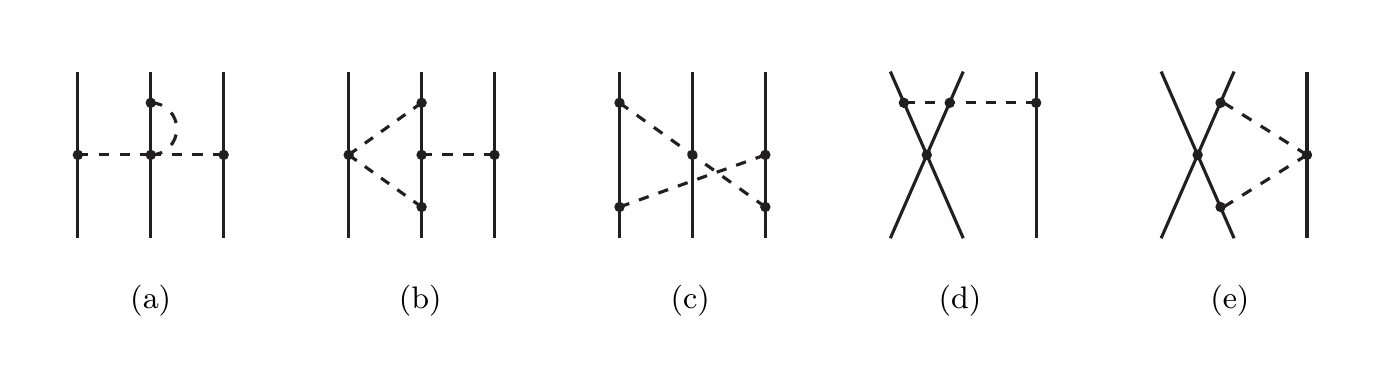}\hspace{0.01in} 
\vspace*{-0.5cm}
 \caption{ Some diagrams of the subleading 3NF, each being representative of a particular topology: (a) 2PE; (b) 2P1PE; (c) ring; (d) 1P-contact; (e): 2P-contact. Note that the 1P-contact topology makes a vanishing contribution.
}
\label{3nf_n3lo}
\end{figure*}

Three-nucleon forces first appear at the third order of the chiral expansion (N$^2$LO) of the $\Delta$-less theory, which is the one we apply in this work. At this order, the 3NF consists of three contributions~\cite{Epe02}: the long-range two-pion-exchange (2PE) graph, the medium-range one-pion-exchange (1PE) diagram, and a short-range contact term. For completeness, we show all  topologies in Fig.~\ref{3nf_n2lo}; note, however, that the contributions depending on $c_4$, $c_D$, and $c_E$ vanish in neutron matter~\cite{HS10}.

In infinite matter, these 3NF can be expressed as density-dependent effective two-nucleon interactions as derived in Refs.~\cite{holt09,holt10}. They are represented in  terms of the well-known non-relativistic two-body nuclear force operators and, therefore, can be conveniently incorporated in the usual $NN$ partial wave formalism and the particle-particle ladder approximation for computing the EoS. The effective density-dependent two-nucleon interactions at N$^2$LO consist of six one-loop topologies. Three of them are generated from the 2PE graph of the chiral 3NF and depend on the LECs $c_{1,3,4}$, which are already present in the 2PE part of the $NN$ interaction. Two one-loop diagrams are generated from the 1PE diagram, and depend on the low-energy constant $c_D$. Finally, there is the one-loop diagram that involves the 3NF contact diagram, with LEC $c_E$. Again, the last two sets do not contribute in neutron matter.

 The 3NF at N$^3$LO has been derived~\cite{Ber08,Ber11} and applied in some nuclear many-body systems~\cite{Tew13,Dri16,DHS19,Heb15a}. The long-range part of 
 the subleading chiral 3NF consists of (cf. Fig.~\ref{3nf_n3lo}): the 2PE topology, which is the longest-range component of the subleading 3NF, the two-pion-one-pion exchange (2P1PE) topology, and the ring topology, generated by a circulating pion which is absorbed and reemitted from each of the three nucleons. 
The in-medium $NN$ potentials corresponding to these long-range subleading 3NFs are given in Ref.~\cite{Kais19} for SNM and in Ref.~\cite{Kais20} for NM. The short-range subleading 3NF consists of (cf. Fig.~\ref{3nf_n3lo}): the one-pion-exchange-contact topology (1P-contact), which gives no net contribution, the two-pion-exchange-contact topology (2P-contact), and relativistic corrections, which depend on the $C_S$ and the $C_T$ LECs of the 2NF and are proportional to $1/M$, where $M$ is the nucleon mass. We include those contributions as well and find them to be in the order of a fraction of 1 MeV.
The in-medium $NN$ potentials corresponding to the short-range subleading 3NFs can be found in Ref.~\cite{Kais18} for SNM and in Ref.~\cite{Treur} for NM.

The LECs we use in this work are displayed in Table~\ref{lecs_I} (for completeness,we also list $c_4$, even though it is redundant for NM. A technical remark is in place: speaking in practical terms, when the subleading 3NFs are included, the $c_1$ and $c_3$ LECs are replaced by -1.20 GeV$^{-1}$ and -4.43 GeV$^{-1}$, respectively. This is because the subleading two-pion-exchange 3NF has the same mathematical structure as the leading one and, thus, and the subleading two-pion-exchange 3NF can be accounted for 
with a shift of the LECs equal to -0.13 GeV$^{-1}$ (for $c_1$) and  0.89 GeV$^{-1}$ (for $c_3$)~\cite{Ber08}.

\begin{figure*}[!t] 
\centering
\hspace*{-1cm}
\includegraphics[width=6.7cm]{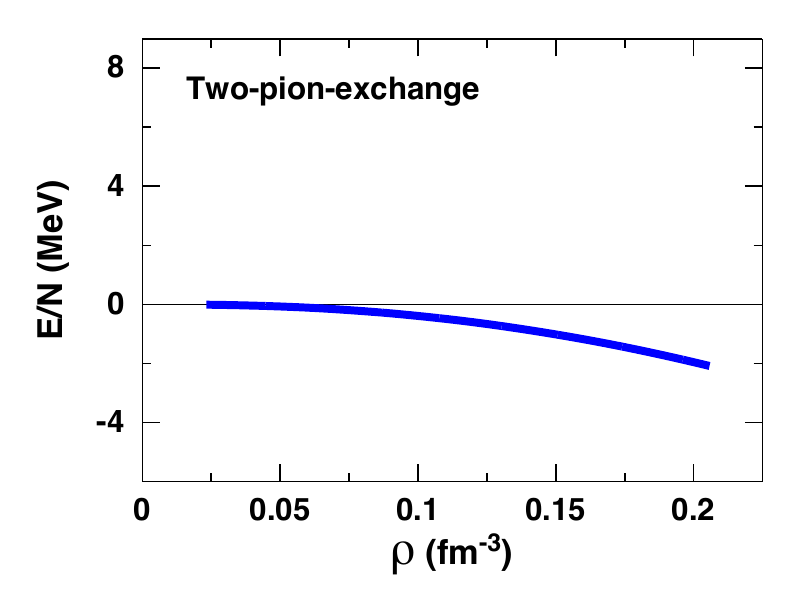}\hspace{0.01in}
 \includegraphics[width=6.7cm]{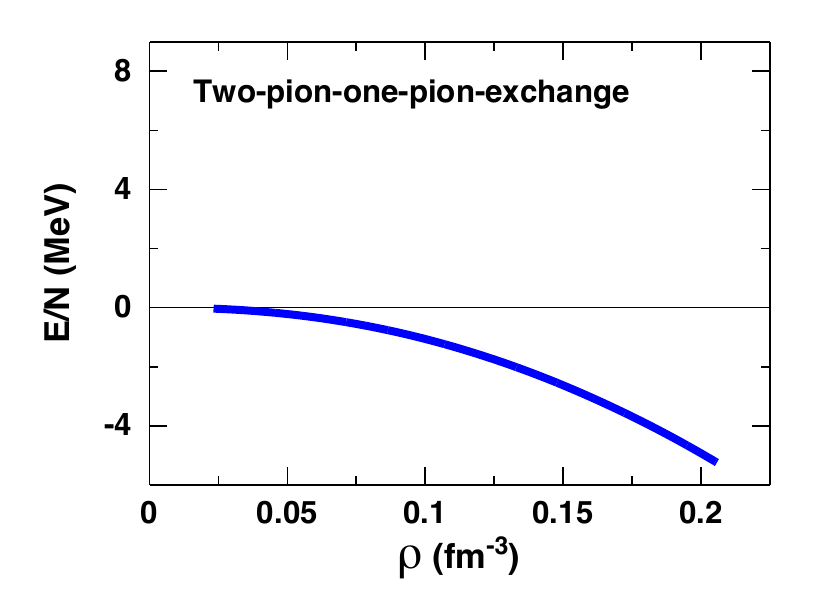}\hspace{0.01in} 
\includegraphics[width=6.7cm]{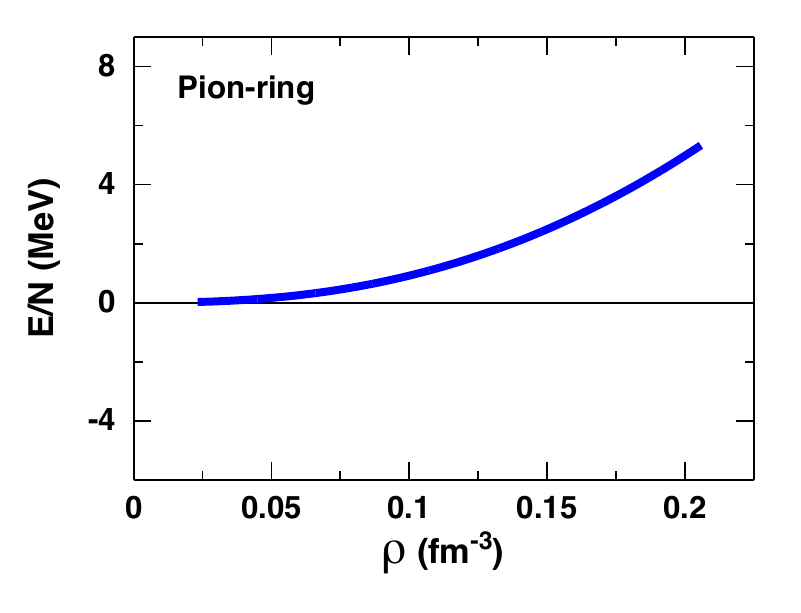}\hspace{0.01in}
 \includegraphics[width=6.7cm]{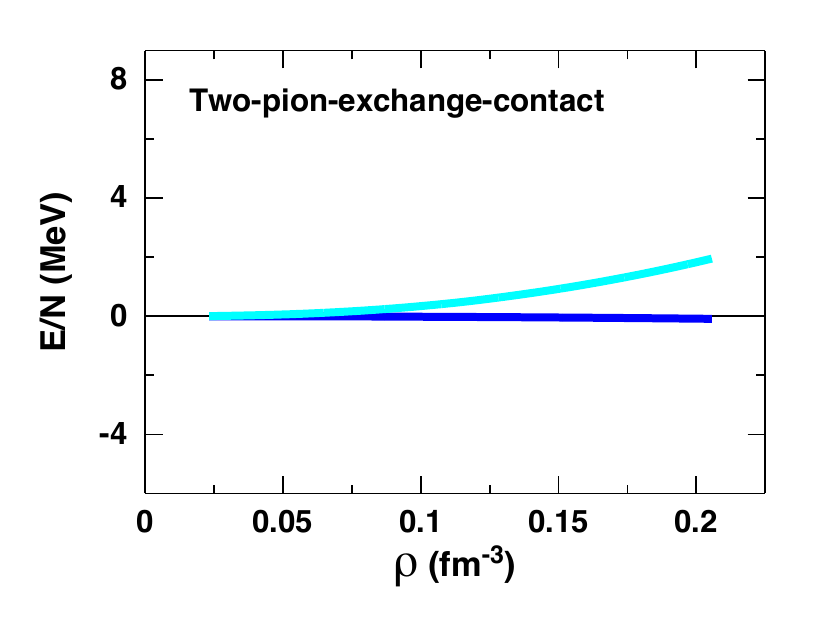}\hspace{0.01in} 
\vspace*{-0.5cm}
 \caption{Various contributions to the 3NF at N$^3$LO in Born approximation. In each case, the topology is indicated inside the frame. For the two-pion-exchange-contact term, which depend sensitively on the chosen 2NF, we show our current predictions (blue) and those we obtain using EM500~\cite{EM03} (cyan) instead of the N$^3$LO potential from Ref.~\cite{EMN17}.
}
\label{3nf_v}
\end{figure*}   

Our results, displayed in Fig.~\ref{3nf_v}, are in excellent agreement with the calculations of Ref.~\cite{Tew13}. This is an important point, because the methods of applying the 3NFs in NM calculations are different in Ref.~\cite{Tew13} as compared to our method. 
 It is worth noting, though, that the two-pion-exchange-contact term is extremely sensitive to the chosen 2NF, through $C_S$ and $C_T$, as demonstrated in Ref.~\cite{Tew13}. For this case, we show our current predictions (blue curve) and those we obtain using EM500~\cite{EM03} (cyan curve). Values obtained with the EGM 450/700~\cite{EGM} are of very similar size as those represented by the cyan curve in Fig.~\ref{3nf_v} but opposite in sign~\cite{Tew13}. 

Overall, Fig.~\ref{3nf_v} indicates that the net impact of the subleading 3NF is small and attractive, due to cancellations between the more sizable contributions.

\subsection{The symmetry energy} 
\label{iso}

The EoS of isospin asymmetric matter, $e(\rho,\alpha)$, is typically expressed as a series expansion with respect to the isospin asymmetry parameter, $\alpha = \frac{\rho_n - \rho_p}{\rho_n + \rho_p}$, where $\rho_n$ and $\rho_p$ are the neutron and proton densities, respectively:
\begin{equation}
e(\rho,\alpha) = e(\rho, \alpha=0) + \frac{1}{2} \Big ( \frac{\partial^2 e(\rho,\alpha)}{\partial \alpha^2} \Big )_{(\alpha = 0)}  \alpha^2 + \mathcal{O} (\alpha^{4})  \; .
\label{e_exp}
\end{equation} 
 Neglecting terms of order $\mathcal{O} (\alpha^{4})$, Eq.~(\ref{e_exp}) takes the parabolic form:
\begin{equation}
e(\rho,\alpha) \approx e_0 + e_{sym} \ \alpha^2  \; , 
\label{asym_e}
\end{equation}
where  $e_{\text{sym}}$ = 
  $\frac{1}{2} \Big ( \frac{\partial^2 e(\rho,\alpha)}{\partial \alpha^2} \Big )_{\alpha = 0}$.
If $\alpha = 1$, the symmetry energy (in the parabolic approximation) becomes the difference between the energy per particle in NM and the one in SNM:
\begin{equation}
e_{sym} (\rho) = e_1 (\rho) - e_0 (\rho) \; .
\label{xxx}
\end{equation}

The expansion of the symmetry energy with respect to density about the saturation point helps identifying useful parameters:
\begin{equation}
\label{yyy}
e_{sym} (\rho) \ \approx \ e_{sym} (\rho_{o}) + L \ \frac{\rho -\rho_{o}}{3 \rho_o} + \frac{K}{2} \frac{(\rho - \rho_{o})^2}{(3\rho_o)^2}  \; .
\end{equation}
 $L$ is referred to as the slope parameter, and is a measure of the slope of the symmetry energy at saturation:
\begin{equation}
\label{L}
L=3\rho_{o} \Big( \frac{\partial e_{sym}(\rho)}{\partial \rho} \Big)_{\rho_{o}}  \; .
\end{equation}
Furthermore, it is obvious from Eqs. (\ref{xxx}) and (\ref{L}) that $L$ is a measure of the slope of the NM EoS at saturation density since the SNM EoS has a vanishing slope at that point, by definition of saturation.
From a variety of phenomenological models, a typical range for $L$ can be stated as (70 $\pm$ 20) MeV \cite{BB_1_2016, BCZ_1_2019, ADS_1_2012, MMSY_1_2012, MSIS_1_2016}.

The parameter $K$  characterizes the curvature of the symmetry energy at saturation density:
\begin{equation}
K=9 \ \rho_{o}^2 \Big( \frac{\partial^2 e_{sym}(\rho)}{\partial \rho^2} \Big)_{\rho_{o}}  \; .
\label{ksym}
\end{equation}

For the EoS of SNM at any order, we use an empirical parametrization as in Ref.~\cite{Oya2010}, where the energy per nucleon is -16.0 MeV at the saturation density $\rho_o$=0.155 fm$^{-3}$. This is to single out the behavior of the energy and pressure in neutron matter and their impact on the symmetry energy while the isoscalar properties remain unchanged -- varying simultaneously isoscalar properties would obscure what we wish to highlight.

\section{Results and discussion }
\label{res}

\subsection{The energy per neutron} 

In our previous calculations~\cite{arx2018}, we either included only the leading 3NF, or, in addition, we accounted only for the subleading 2PE 3NF. With the inclusion of all the subleading 3NF contributions we are now in the position to conduct complete calculations at N$^3$LO and draw robust conclusions.

As pointed out in Sec.~\ref{II}, errors in the $\pi N$ LECs are no longer an issue with regard to uncertainty quantification. On the other hand, central to chiral EFT 
is the truncation error, which we will now address. If observable $X$ is known at order $n$ and at order $n+1$, a reasonable estimate of the truncation error at order $n$ can be expressed as the difference between the value at order $n$ and the one at the next order:
\begin{equation}
\Delta X_n = |X_{n+1} - X_n| \; ,
\label{del} 
\end{equation} 
since this is a measure for what has been neglected at order $n$.

To estimate the uncertainty at the highest order that we consider, we follow the prescription of Ref.~\cite{Epel15}. For an observable $X$ that depends on the typical momentum of the system under consideration, $p$, one defines $Q$ as the largest between $\frac{p}{\Lambda_b}$ and $\frac{m_{\pi}}{\Lambda_b}$, where $\Lambda_b$ is the breakdown scale of the chiral EFT, for which we assume 600 MeV~\cite{Epel15}. The uncertainty of the value of $X$ at N$^3$LO is then given by:
\begin{equation}
\Delta X = \max \{Q^5|X_{LO}|, Q^3|X_{LO} - X_{NLO}|,Q^2|X_{NLO} - X_{N^2LO}|, Q|X_{N^2LO} - X_{N^3LO}| \} \; . 
\label{err}
\end{equation} 
We identify $p$ with the neutron Fermi momentum at the density under consideration. For the cutoff $\Lambda$ of the regulator, Eq.~(\ref{reg}), we use 450 MeV.

\begin{figure*}[!t] 
\centering
\hspace*{-3cm}
\includegraphics[width=7.7cm]{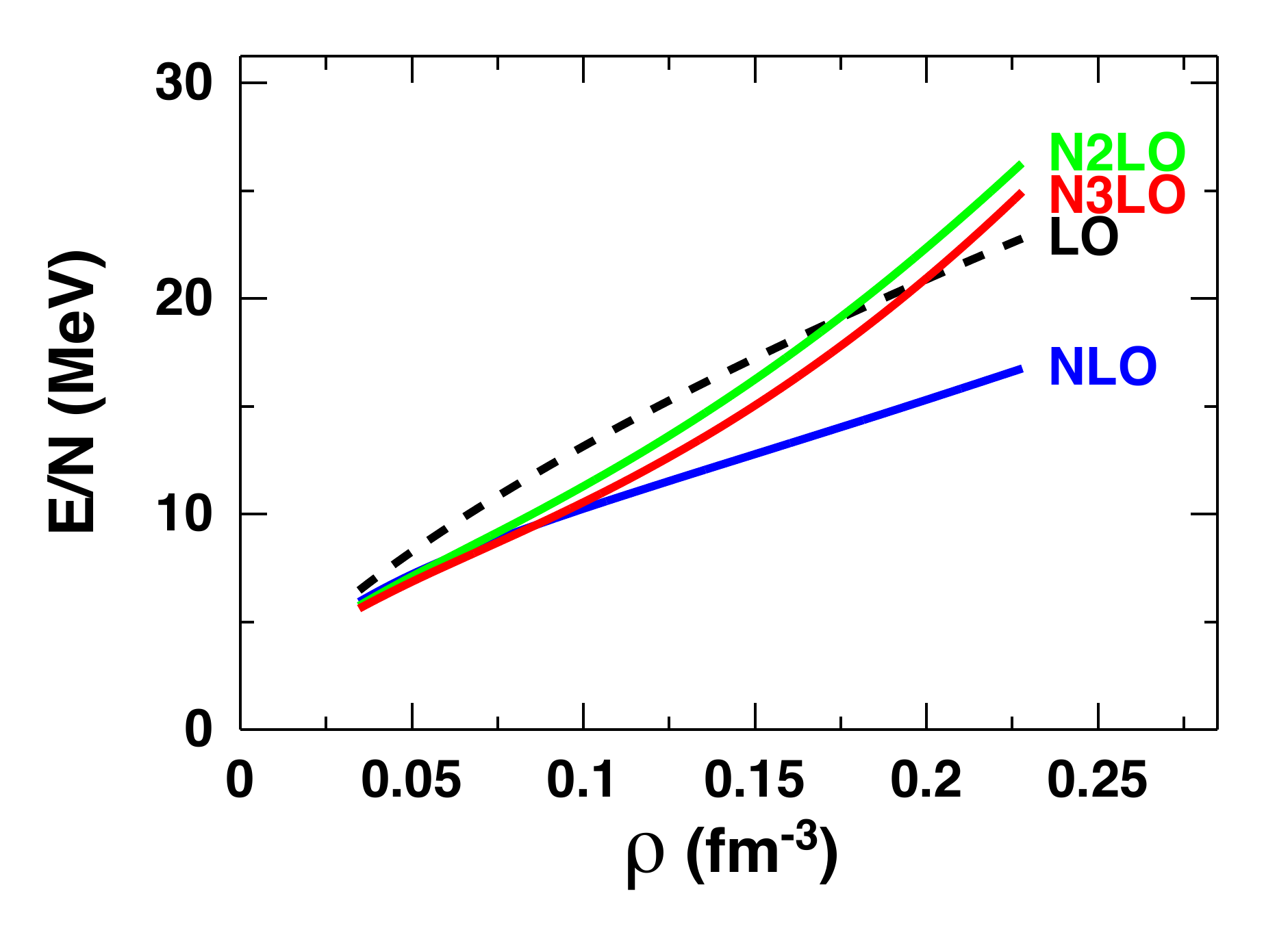}\hspace{0.01in} 
\vspace*{-0.5cm}
 \caption{(Color online) Energy per neutron in NM as a function of density, from leading order (black dash) to fourth order (solid red).
}
\label{nm_eos}
\end{figure*}

In Fig.~\ref{nm_eos}, we show our results for the energy per neutron in NM as a function of density over four orders, from LO to N$^3$LO. We see
large variations from leading order to the next, as to be expected at the lowest orders. It is interesting to notice the very large differences between NLO and N$^2$LO, mostly due to the first appearance of 3NFs. The predictions at N$^3$LO are slightly more attractive than those at N$^2$LO, in agreement with other calculations~\cite{Dri+16}.

\begin{figure*}[!t] 
\centering
\hspace*{-3cm}
\includegraphics[width=7.7cm]{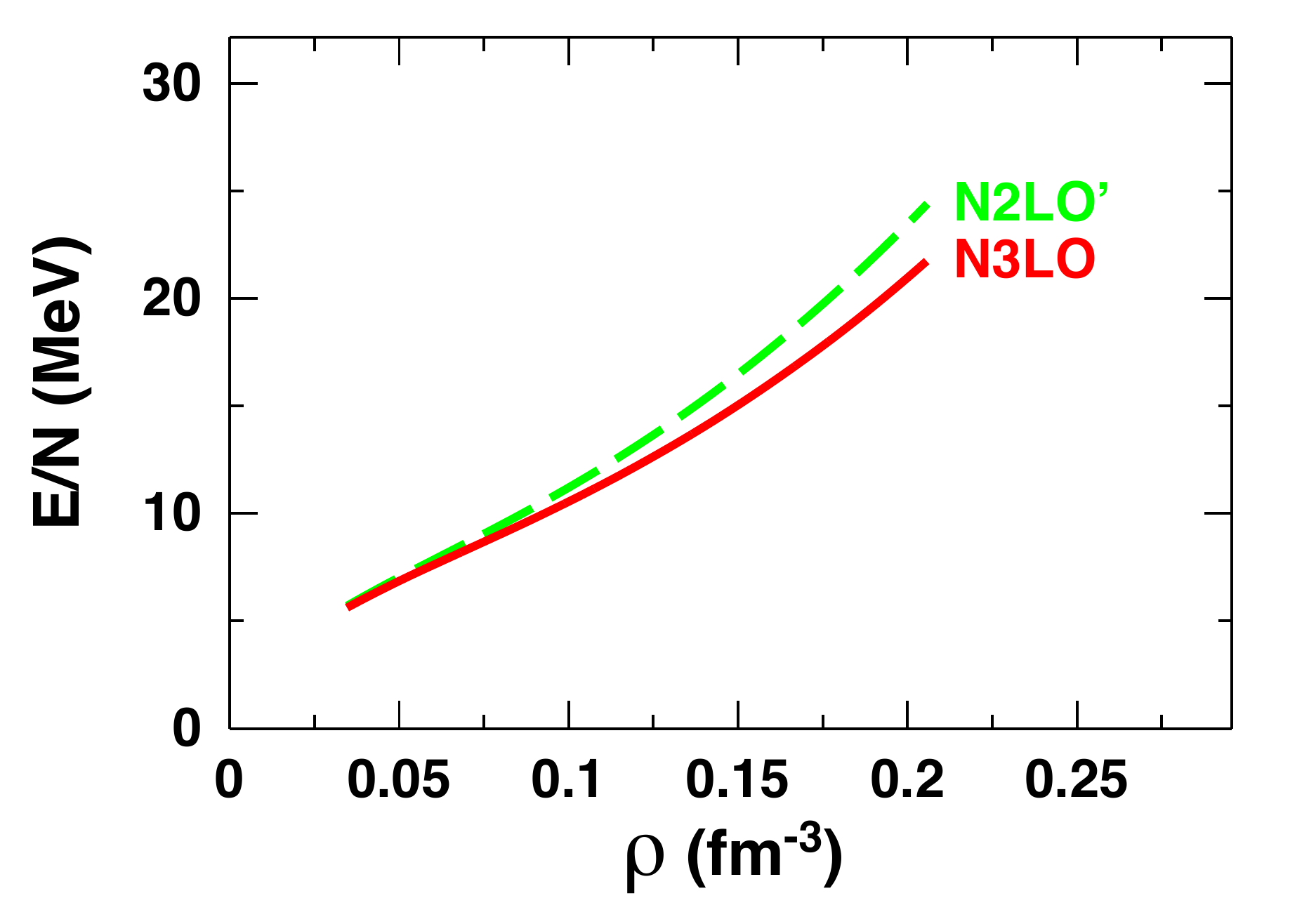}\hspace{0.01in} 
\vspace*{-0.5cm}
 \caption{(Color online) Energy per neutron in NM as a function of density. The solid (red) curve is the same as in Fig.~\ref{nm_eos}, while in the dashed green line (denoted by N$^2$LO'), 
the N$^3$LO 3NF contribution has been left out.}
\label{nm23}
\end{figure*}

Figure~\ref{nm23} shows the impact of the subleading 3NF contribution only, comparing the result of the complete calculation at N$^3$LO with the one obtained with the 2NF at N$^3$LO and the leading 3NF only,  model (N$^2$LO$'$). The effect is mildly attractive, about 2 MeV at the highest density, mostly due to the subleading 2PE 3NF. 

The first four-nucleon forces (4NFs) appear at N$^3$LO, but were found to be negligible~\cite{Tew13}. Therefore, we omit 4NFs.

 Our highest-order results for the energy per neutron at different densities are shown in the second column of Table~\ref{par0} with their chiral uncertainty calculated as in Eq.~(\ref{err}). The choice of these densities is motivated in the next section.

Our NM EoS is rather soft within the large spectrum of theoretical predictions in the literature, which is generally true for predictions based on chiral EFT. The degree of softness is best discussed in the context of density dependence of the symmetry energy, see next subsection.

\subsection{The symmetry energy: microscopic predictions and empirical constraints}

\begin{figure*}[!t] 
\centering
\hspace*{-3cm}
\includegraphics[width=7.7cm]{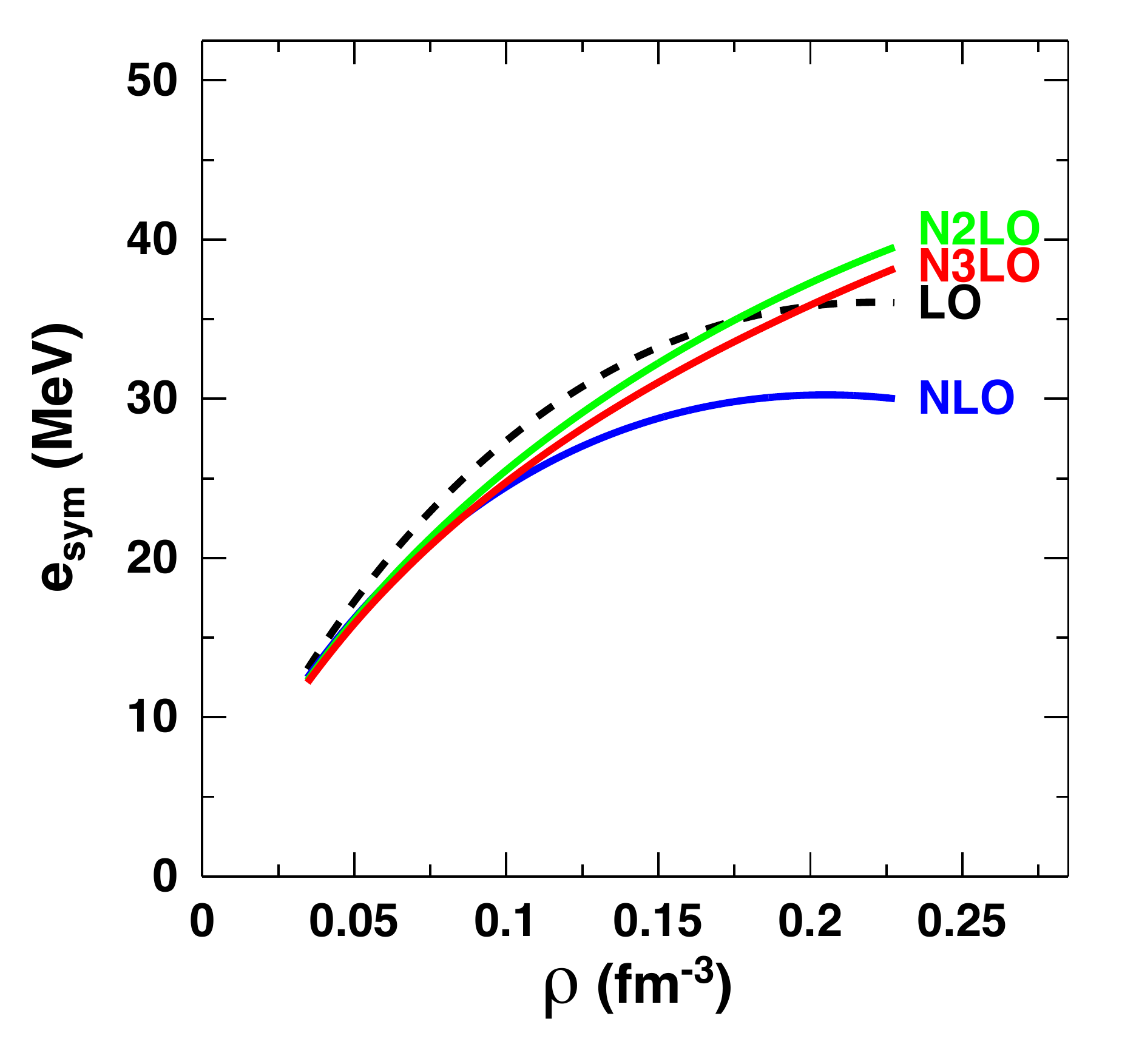}\hspace{0.01in} 
\vspace*{-0.5cm}
 \caption{(Color online)  The symmmetry energy as a function of density, from leading order (black dash) to fourth order (solid red).
}
\label{esym}
\end{figure*}

Our order-by-order predictions for the symmetry energy are displayed in Fig.~\ref{esym}, and our results at N$^3$LO together with their truncation errors are given in Table~\ref{par0}.
To understand the full implications of our predictions, several comments are in place.
Correlations between physical observables and the symmetry energy or its first derivative have been explored extensively, mostly using families of phenomenological models. Popular examples are the Skyrme forces~\cite{B_1_2000} or relativistic mean-field models (RMF) \cite{PF_1_2019}. These models are parameterized so as to ensure that the empirical saturation properties are well described, while the models can differ considerably in the isovector properties. Earlier investigations with a family of Skyrme interactions concluded that there is a linear correlation between the slope parameter and the neutron skin thickness of $^{208}$Pb \cite{B_1_2000}. This inherent connection between the symmetry energy density derivative and the neutron skin of neutron-rich nuclei is of great interest, because accurate measurements of the skin should then allow to set stringent constraints on the density dependence of the symmetry energy around saturation.

Relativistic mean-field models predict a very wide range of $L$ values, for example IU-FSU \cite{FHPS_1_2010} gives 47.2 MeV for $L$, while NL3 \cite{LKR_1_1997} yields a value of 118.2 MeV.  Naturally, these models also produce a large range of neutron skin values.  For neutron skin predictions and RMF models, see also Ref.~\cite{MCVW_1_2011}, where the authors utilize a large set of RMF models constrained by accurate fits of the nuclear binding energies and charge radii.

 Additional studies that have explored these correlations, using a variety of phenomenological and theoretical models, are provided in Refs.~\cite{SDLD_1_2015, MADSCV_1_2017, TLOK17, MADS_1_2018, TRRSWM_1_2018, HY_1_2018}. Constrains on $L$ vary considerably depending on the methods employed~ \cite{ADS_1_2012, ADSCS_1_2013, VCRW_1_2014}.  Further analyses that employ laboratory data to extract constraints on the density dependence of the symmetry energy can be found in Refs.~\cite{Tsang+09, Tsang+12, LL13, Kort+10, DL_1_2014, Roca+15, Tam+11, Brown13, R_etal_1_2011, R_etal_1_2016}. As for the symmetry energy curvature, Eq.~(\ref{kappa}), constraints on $K$ have large uncertainty~\cite{VPPR_1_2009, DMPV_1_2010, SDLD_1_2014}.

In addition to the energy per neutron at saturation density, we show in Table~\ref{par0} the symmetry energy at saturation, the slope parameter as defined in Eq.~(\ref{L}), and the pressure in neutron matter. As mentioned earlier, 
a softer nature is typical of chiral predictions, see, for instance, Ref.~\cite{chi}, where SRG-evolved interactions based on the potentials from Ref.~\cite{EM03} and the leading 3NF are employed. A comparison with phenomenological interactions of the past, such as Argonne V18 and the UIX 3NF~\cite{APR}, is given in Ref.~\cite{chi}.
For a more recent analysis, see Ref.~\cite{Bays20}, where the reported values for $e_{sym}(\rho_o)$ and $L$  are (31.7 $\pm$ 1.1) MeV and (59.8 $\pm$ 4.1) MeV, respectively.

\begin{figure*}[!t] 
\centering
\hspace*{-3cm}
\includegraphics[width=7.0cm]{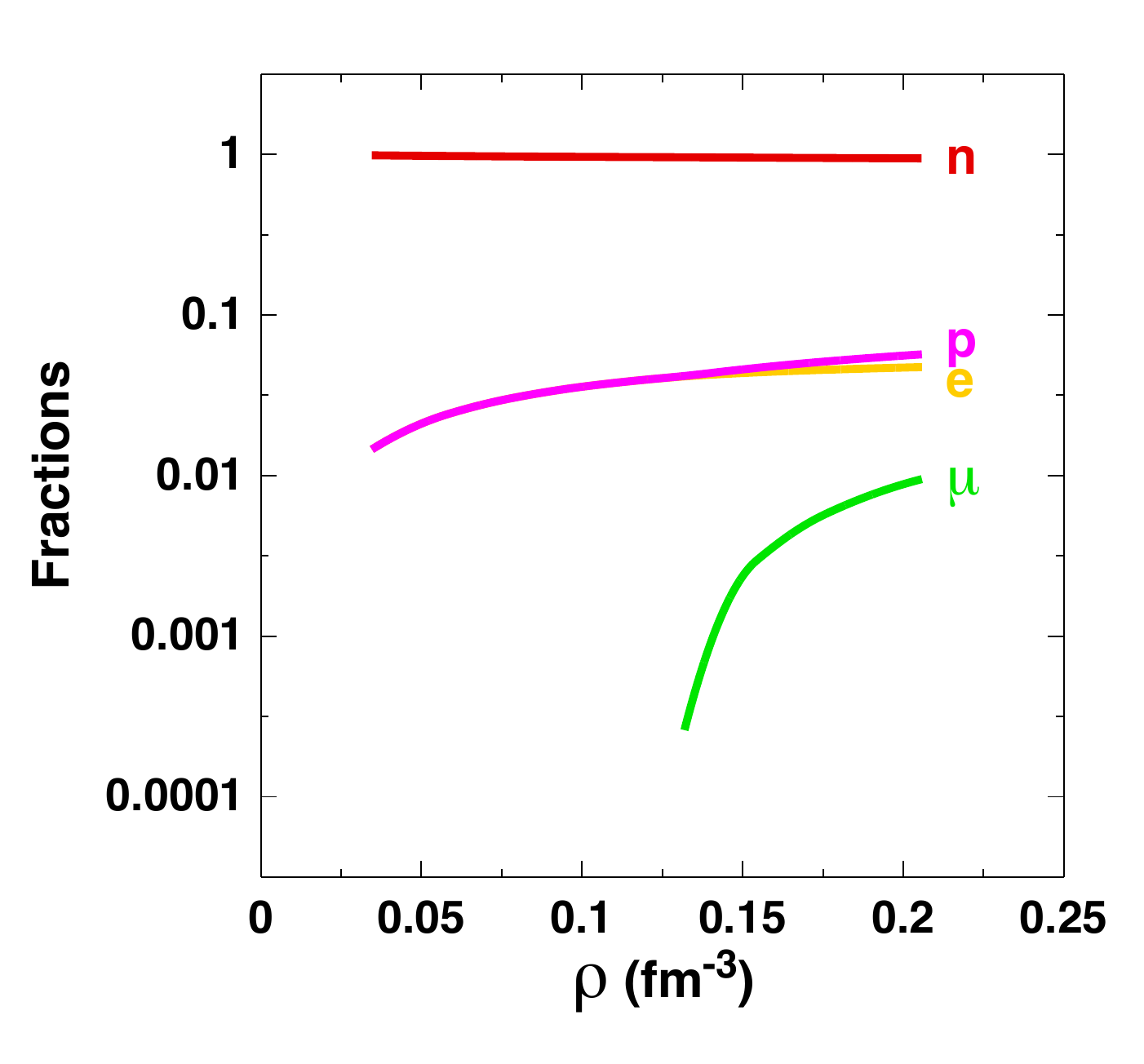}\hspace{0.01in} 
\vspace*{-0.5cm}
 \caption{(Color online) Particle fractions as a function of density in $\beta$-equilibrated matter with neutrons, protons, electrons, and muons.
}
\label{frac}
\end{figure*}

On the other hand, values such as those shown in the first row of Table~\ref{par0} -- approximately $L$=(50 $\pm$ 10) MeV, and pressure at $\rho_o$ between 2 and 3 MeV/fm$^3$--   are nowhere near those extracted from the PREX-II experiment~\cite{prexII}, which are: (38.29 $\pm$ 4.66) MeV and (109.56 $\pm$ 36.41) MeV, for $e_{sym}$ and $L$, respectively. The corresponding value of the pressure at $\rho_o$ is then, approximately, between 3.66 MeV/fm$^3$ and 7.30 MeV/fm$^3$. Furthermore, such stiff symmetry energy would allow rapid cooling through direct Urca processes to proceed at unusually low values of the neutron star mass and central density~\cite{prexII}, which seems unlikely~\cite{PLPS09}.  The various particle fractions in $\beta$-stable matter we obtain with our N$^3$LO predictions are shown in Fig.~\ref{frac}. The proton fraction is close to 6\% at $\rho \approx$ 0.2 fm$^{-3}$, still far from the direct Urca threshold of approximately 11\%.

Back to Table~\ref{par0}, we also show the predictions at some specific densities below $\rho_o$. These are the  densities identified in Ref.~\cite{Tsang} as ``sensitive" densities from the slope of the correlation in the plane of $e_{sym}(\rho_o)$  {\it vs.} $L$ obtained from the measurements of a specific observable. In fact, a particular slope reflects a specific density at which that observable is especially sensitive to the symmetry energy. Our ab initio predictions and the values taken from Ref.~\cite{Tsang} shown in Table~\ref{par0}, compare favorably within uncertainties.
 We recall that, at $\rho$ = (2/3)$\rho_o \approx$ 0.1 fm$^{-3}$ (an average between central and surface densities in nuclei),  the symmetry energy is well constrained by the binding energy of heavy, neutron-rich nuclei~\cite{prexII} -- hence, the relevance of this density region for the purpose of correlations between $L$ and the neutron skin of $^{208}$Pb.

Finally, in Table~\ref{kappa} we show our predictions for $K$ as defined in Eq.~(\ref{ksym}) in comparison with recent constraints~\cite{Tsang}. Given the strong sensitivity of $K$ to the details of the symmetry energy curvature, this parameter is very model dependent and difficult to constrain, with reported values ranging from large and negative to large and positive. As seen from Table~\ref{kappa}, our predictions are in reasonable agreement with the constraints discussed in Ref.~\cite{Tsang}.

 It is our understanding that, focusing on the sensitive density for a given observable, consistency among different analyses can be found~\cite{Tsang}.
Perhaps these considerations may help with the interpretation of the large values from PREX-II. When the current constraint from PREX-II is included in the fits, its impact is weak due to the large experimental uncertainty~\cite{Tsang}.  For the slope of the symmetry energy at $\rho$ = (2/3)$\rho_o \approx$ 0.1fm$^{-3}$, the value obtained from PREX-II data~\cite{prexII} is  $L$ =(73.69 $\pm$ 22.28) MeV.

\begin{table*}
\caption{The energy per neutron, the symmetry energy, the slope parameter, and the pressure at N$^3$LO at various densities, $\rho$, in units of $\rho_o$=0.155 fm$^{-3}$.
$L$ is defined as in Eq.~(\ref{L}) at the specified density. The values in parentheses are taken from Ref.~\cite{Tsang}. The constraint for $L$ in the third row ($\rho =0.67 \rho_o$) is given for  $\rho$=0.1 fm$^{-3}$. The constraint at $\rho =0.31 \rho_o$ is from Ref.~\cite{ZZ15}. }
\label{par0}
\centering
\begin{tabular*}{\textwidth}{@{\extracolsep{\fill}}ccccc}
\hline
\hline
 $\rho$ $ (\rho_o) $ & $ \frac{E}{N}(\rho)$ (MeV) & $e_{sym}(\rho)$ (MeV)  & $L (\rho)$(MeV)  &  $P_{NM}(\rho)$ (MeV/fm$^3$)  \\
\hline
\hline
 1 & 15.56  $\pm$ 1.10& 31.57 $\pm$ 1.53 (33.3 $\pm$ 1.3)  & 49.58 $\pm$ 8.47 (59.6 $\pm$ 22.1) & 2.562  $\pm$ 0.438 (3.2 $\pm$ 1.2)  \\  
0.72 (0.72 $\pm$ 0.01) & 11.52  $\pm$ 0.43 & 26.46 $\pm$ 0.82 (25.4 $\pm$ 1.1) & 44.91 $\pm$ 3.40 & 1.05  $\pm$ 0.13  \\  
 0.67 (0.66 $\pm$ 0.04) & 10.81  $\pm$ 0.41 & 25.25 $\pm$ 0.72 (25.5 $\pm$ 1.1) & 44.65 $\pm$ 3.23 (53.1 $\pm$ 6.1) & 0.859  $\pm$ 0.120  \\  
0.63 (0.63 $\pm$ 0.03) & 10.39  $\pm$ 0.41 & 24.47  $\pm$ 0.66 (24.7 $\pm$ 0.8) & 43.81 $\pm$ 3.11 & 0.748  $\pm$ 0.116  \\  
0.31 (0.31 $\pm$ 0.03) &   6.715 $\pm$ 0.086  &15.43 $\pm$ 0.12  (15.9 $\pm$ 1.0) &  32.35 $\pm$ 0.55    &     0.174 $\pm$ 0.008                            \\ 
0.21 (0.22 $\pm$ 0.07) & 5.472  $\pm$  0.039 & 11.73 $\pm$ 0.05 (10.1 $\pm$ 1.0) & 27.57 $\pm$ 0.11 & 0.106  $\pm$ 0.002  \\  
\hline
\hline
\end{tabular*}
\end{table*}

 \begin{table*}
\caption{Predictions at N$^3$LO and constraints for the parameter $K$ at two densities. }
\label{kappa}
\centering
\begin{tabular*}{\textwidth}{@{\extracolsep{\fill}}ccc}
\hline
\hline
 $\rho$ (fm$^{-3}$) & $ K(\rho)$ (MeV) & $K(\rho)$ (MeV) from Ref.~\cite{Tsang}  \\
\hline
\hline
 0.155 & -101   $\pm$ 23 & -180  $\pm$ 96   \\   
  0.1 & -68.7   $\pm$ 15.7 & -79.2  $\pm$ 37.6   \\   
\hline
\hline
\end{tabular*}
\end{table*}

\section{Conclusions and outlook}                                                                  
\label{Concl} 

We presented results of the energy per neutron in NM based on recent high-quality chiral potentials and 3NFs up to N$^3$LO. Subleading 3NFs do not bring in new free parameters, but some short-range contributions and relativistic corrections do depend on the choice of the $NN$ potential through the LECs of the 2NF. Our order-by-order calculations show systematic improvements of the predictions and allow for estimating the chiral uncertainty.

Although results are well converged at N$^3$LO, specific contributions of the 3NF at N$^4$LO may play a role -- note that the contact terms of the 3NF at N$^4$LO seem promising towards the solution of some long-standing problems in low-energy few-body reactions~\cite{Ay}. However, complete calculations at N$^4$LO are necessary for reliable conclusions concerning the fifth order, and, realistically, those will not be available in the near future. 

Our NM EoS is rather soft on the scale of theoretical predictions available in the literature. We analyzed this aspect through the density dependence of the symmetry energy. 
 We compared our current predictions with recent constraints, extracted with methods that are different than those typically used to narrow down the symmetry energy and its slope at saturation~\cite{Tsang}. We found good agreement with the provided values of the symmetry energy at and below saturation, as well as the slope and the curvature of the symmetry energy at saturation. On the other hand, we noted that the generally soft predictions based on chiral EFT are far from the values of $e_{sym}$ and $L$ obtained from PREX-II~\cite{prexII}.

 Exploiting the sensitivity of selected observables to the isovector part of the EoS at some densities, the symmetry energy can be constrained through the measurement of such observables.
In the past, we have expressed concerns (see, for instance, Ref.~\cite{ASY}), about some aspects of the methodologies utilized to extract symmetry energy constraints around saturation density from measured observables, in particular about the degree of model dependence of the constraints, which do not provide a functional relation between density and symmetry energy. Hopefully, a different approach to the analyses  will reduce the model dependence and the discrepancies among the constraints given at saturation density.

\section*{Acknowledgments}
This work was supported by 
the U.S. Department of Energy, Office of Science, Office of Basic Energy Sciences, under Award Number DE-FG02-03ER41270. 

 
\end{document}